\documentclass[conference]{IEEEtran}
\usepackage{cite}

\ifCLASSINFOpdf
  \usepackage[pdftex]{graphicx}
   \graphicspath{}
  \DeclareGraphicsExtensions{.pdf,.jpeg,.png,.eps,.gif,.jpg }
\else
\fi
\usepackage{CJK}
\usepackage[numbers,sort&compress]{natbib}
\usepackage{cite}
\usepackage{amsmath,amssymb,enumerate}
\usepackage{cite,graphics,graphicx}
\usepackage{graphicx}
\usepackage{subfigure}
\usepackage{listings}
\usepackage{makeidx}
\usepackage{hyperref}
\usepackage{algorithm}
\usepackage{algorithmic}
\usepackage{float}
\usepackage{multirow}
\usepackage{amsmath}
\usepackage{amssymb}
\usepackage{xcolor}
\usepackage{epsfig}
\usepackage{graphics}
\usepackage{listings}
\usepackage{algorithm}
\usepackage{algorithmic}
\hypersetup{linkcolor=black, %
            citecolor=black, %
             pdftitle={Title}, %
            pdfauthor={Name}, %
           pdfsubject={Subject}, %
          pdfkeywords={Key words}}
\begin{document}
%
\title {  Solving The Longest Overlap Region Problem for Noncoding DNA Sequences with GPU}

\author{\IEEEauthorblockN{YuKun Zhong\IEEEauthorrefmark{1},
JianBiao Lin\IEEEauthorrefmark{2},
BaoQiu Wang\IEEEauthorrefmark{2},
Chen Tao\IEEEauthorrefmark{3},
Che Nian\IEEEauthorrefmark{4},
Xie Wen\IEEEauthorrefmark{4}
}
\IEEEauthorblockA{
Computer Science and Engineering Department\\ Sichuan University Jinjiang College, Penshan 620860, China\\
Email: zystephen7@gmail.com}
}


%


\maketitle

\begin{abstract}
Early hardware limitations of GPU (lack of synchronization primitives and limited memory caching
mechanisms) can
make
GPU-based
computation
inefficient. Now Bio-technologies bring more chances to Bioinformatics and Biological Engineering. Our paper introduces a way to solve the longest overlap region of non-coding DNA sequences on using the Compute Unified Device
Architecture (CUDA) platform Intel(R) Core(TM) i3- 3110m
quad-core. Compared
to standard CPU implementation, CUDA performance proves the method of the
longest overlap region recognition of noncoding DNA is an efficient approach to high-performance
bioinformatics applications. Studies show the fact that efficiency of GPU performance is more than 20 times speedup than that of CPU serial
implementation. We believe our method gives a cost-efficient solution to the bioinformatics community for solving longest overlap region recognition problem and other related fields.\\\\
keywords: CUDA, GPUs, RMQ, suffix array, DC3,
LCP, Noncoding DNA,
\end{abstract}


%
\IEEEpeerreviewmaketitle

\section{Introduction}
 In the early years, many science researcher held that a large amount of non-coding DNA had not biological functions and viewed non-coding DNA as ”junk DNA”. But, recent studies show many non-coding DNA are functional and beneficial to human beings. For example, many biological scientists found the RNA sequences of telomeres have special function which was ignored by previous research. Sequence alignment is an important problem in computational biology and sequence comparison is an important tool for researchers in molecular biology \cite{bernaola2000finding}. Computational recognition of genes is one of challenges in the analysis of newly sequenced genomes filed,
which is fundamental for modern functional genomics \cite{shi2009accelerating}.\\\\

 In recent years, modern multi-core and many-core architectures
are revolutionary high-performance computing (HPC).
Now CPU microprocessors, based on a single central processing,
promote the performance of computer application to make floating point arithmetic achieved 11 times
per second single chip, the era of many-core processor has
begun \cite{nvidia2007compute}. The emergence of many-core architectures, such as compute
unified device architecture (CUDA)-enabled GPUs and
other accelerator technologies, these emerging technologies open more opportunities to optimize many
biological algorithms and to provide industries with more advanced and powerful computing
hardware. \\\\
Due to exponentially growing of DNA bases and awareness of function of non-coding DNA, computational recognition of genes became a time-consuming and challenging job. The development of bioinformatics need to have a better performance. Several efforts have been  made to optimize the biological algorithm and reach more accurate results. Life science have emerged as a primary application area
for use of GPU computing. Now a method to recognize the longest overlap region of DNA is on two different platforms: multi-core( CPUs) and
many-core(GPU). GPU performance grows faster than development of CPU in  many fields of bioinformatics. In this paper, we present a modified and efficient parallel algorithm with CUDA to solve the longest overlap region problem. Based on DC3 algorithm which is used to construct suffix array and RMQ algorithm, we change series programming into computation of the problem in parallel. Here DC3 algorithm is more than 20 times speedup than that of CPU serial implementation and the fact shows solving the longest overlap region with CUDA is a
perfect method.\\\\

\section{general introduction of GPU and CUDA programming model }

  Recently, more and more application developers pay much attention to GPUs. The new products of GPUs are dramatically increasing programmability and generality but still providing developers with huge memory bandwidth and powerful computational ability \cite{liao2013gpu}. In order to meet growing requirements of programming developers, many industries have been engaging into designing processors such like NVIDIA used a large amount of processors cores to built programming processor \cite{ryoo2008optimization}. Culminating in NVIDIA’s
first GPU in 1999. After one year when NVIDIA have been developing GPU terms, many software and computer games were not only filed which made remarkable breakthroughs with technology. The
General Purpose GPU (GPGPU) movement had dawned \cite{wen2011gpu}. \\

CUDA is a parallel computing platform and programming
model invented by NVIDIA. \cite{nvidia2007compute} \cite{wen2011gpu}. Since yielded in 2007, hundreds of millions
of computers equipped with CUDA processing capability
of GPU is widely used. GPU parallel technology
is now widely used in various fields, GPU high-speed
parallel processing capabilities could be used in digital image
processing algorithms, discrete simulation, general computing,
greatly improving computational speed, because GPU performance characteristics can be applied to numerical
calculation and matrix processing \cite{sierra2010parallel}.
 \begin{figure}[]
\centering
\graphicspath{}
\includegraphics[width=2.5in]{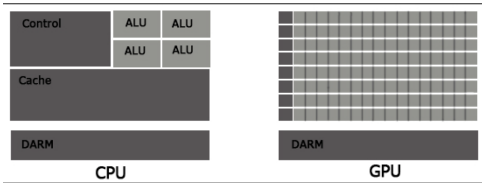}
 \DeclareGraphicsExtensions.
\caption{GPU CPU architecture}
\label{fig_18}
\end{figure}
\\\\

 CUDA programming model is very well suited
to parallel capabilities of GPUs. CUDA devices collect a plenty of data parallelism to accelerate data process. Two control circuits
are relatively simple to GPU, and demand for Cache is small,
so you can put most of transistors in the cell.
$40{\%}$ of GPU is ALU. Illustrate in Figure.\ref{fig_18} .\\\\

 It is a common CUDA programming model that sequential host program processes organize an application and parallel device can process at least one host programme. Typically, parts of host programme are implemented on CPU and GPU take responsibility of the others parallelized parts. Threads are rationally divided into each thread block by programmer; kernel that we used thus is made up with a grid which consist of at least one thread block. A cluster of threads synchronously work through barrier synchronization in a thread block which shared a same memory space \cite{satish2009designing}.\\\\

\section{LCP and related algorithm}
In this paper, we present the longest overlap region problem that is to compute
the longest common prefix(LCP) non-coding DNA sequences and its Pos array in suffix arrays. Given a reference sequence $S = {a_{1} a_{2} a_{3} a_{4} a_{5}..a_{n}}$. For i $=$ 1 , 2 , 3..n, every S(i,n) is a suffix of S, also every S(1,i) is the prefix of S. For convenience, we define LCP(S,i,j) as the longest common prefix between SA[i,n]\hspace{2mm}(SA is the suffix  array of S) and SA[j,n].\\\\

 Suffix array is a space-efficient data structure that guarantee accurate and quick searching of a subsequence from pattern. Pos of all suffixes of a sequences are stored basically in Suffix array \cite{kasai2001linear}. Compared with suffix trees, We naturally viewed Suffix arrays
as a more efficient and convenient method data structure. This
structure can be served as an array of integers representing
the start position of every lexicographically ordered suffix of
a string. In the table, we can use SA of S or SA{S} to indicate suffix array of S. It is easy to find the rules of suffix array of S that an array SA[0..n] which contains a permutation of the integers 0..n so, we can make a conclusion that
$S[SA[0]..n] < S[SA[1]..n] < .... < S[SA[n]..n]$, meanwhile we can give a rule: SA[m] = k iff S[k..l]
is the $m^{th}$ suffix of S \cite{karkkainen2006linear} \cite{puglisi2008space}. Given example $S = { ATTGCTAC }$ , we can build a sort suffix table of S in the Table 1
\begin{table}[!htbp]\centering\normalsize
\caption{SUFFIX INDEX TABLE OF $S{\lbrace}ATTGCTAC{\rbrace}$}
\begin{tabular}{|l|l|l|l|l|l|l|l|l|}
\hline
\hspace{4mm}S &A&T&T&G&C&T&A&C\\
\hline
  INDEX  &0&1&2&3&4&5&6&7\\
 \hline
\hspace{3mm}SA &6&0&7&4&3&5&2&1\\
\hline
\end{tabular}
\end{table}

Traditional method build suffix array can be in O(n*log(n)). In this paper, we give a linear-time
algorithm called DC3  which is a special case of a another cover sample. in this situation, the sample must meet the requirements:  1 : The sample itself can be sorted efficiently. 2 : Sorted sequence of all suffixed is helped by of the sorted sample. sorting suffixes in DC3 method is beginning at location in a difference cover sample
modulo 3, after the process utilize index to find all suffixes \cite{karkkainen2006linear}. It takes the following 2=3-recursive divide-and-conquer
approach :\\
Definition :\\\\
  A set\hspace{2mm}$ D {\subseteq} {[0,v)} $is difference cover module v if
  \begin{equation}
  \lbrace(i-j)\hspace{1mm} mod\hspace{1mm} v \hspace{1mm}\mathbf{|}\hspace{1mm}i,j{\in}D\rbrace={[0,v]}
\end{equation}
A v-periodic sample C of [0,n] with the period D, that is,
\begin{equation}
C=\lbrace i\in[0,n]\hspace{1mm}\mathbf{|}\hspace{1mm}i\hspace{1mm}mod\hspace{1mm} v\in D\rbrace
\end{equation}
is a difference cover sample if D is a difference cover modulo v.\\\\
For k∈[0,..3),\hspace{3mm} define :
\begin{equation}
  B_{k}=\lbrace i{\in}[0,n]\hspace{1mm}\mathbf{|}\hspace{1mm} i\hspace{1mm} mod\hspace{1mm} 3 = k \rbrace
\end{equation}\\
the set of cover position is $C\hspace{1mm}=\cup_{k{\in}1,2}B_{k}$, $\hat{C}= {[0,3]} \backslash C $
\\\\
Step 1 : \hspace{2mm}Built suffix array of the suffixes starting at position i (i\hspace{1mm}mod $3 \neq 0$) and sort sample suffixed
\\

 So we can see $\hspace{1mm}B_{2}{\cup}B_{1}=C\hspace{1mm}$ to be the set of the sample of positions. We can easily find $C=\lbrace1,2,4,5,7\rbrace $ , $B_{2}=\lbrace2,5\rbrace$and $B_{1}=\lbrace 1,4,7\rbrace $. For $k=1,2$, make a string \\ \\ $R_{k}=[t_{i}t_{i+1}t_{t+2}][t_{i+3}t_{i+4}t_{t+5}]...[t_{maxB_{k}}t_{maxB_{k+1}}t_{maxB_{k+2}}]$\\\\
 whose characters are triples $[t_{i}
t_{i+1}t_{i+2}]$.  Let $R=R_{1}\bigodot R_{2}$ be the concatenation of R1 and R2. $R_{1}=\lbrace(C00),(CTA),(TTG)\rbrace$,$R_{2}=\lbrace(TAC),(TGC)\rbrace$ refer to table 1, and we can get $R=\lbrace(CTA),(TTG),(TGC),(C00),(TAC) \rbrace$. For i $\in$ C, let $rank(S_{i})$ denote rank of $S_{i}$ in the sample set of $S_{C}$. For $i\in B_{0}$,$rank(S_{i})$
 is undefined. illustrate in Table.2 \cite{kasai2001linear}
\begin{table}[!htbp]\centering\normalsize
\caption{RANK OF SORTED SAMPLE SUFFIX}
\begin{tabular}{|l|l|l|l|l|l|l|l|l|}
\hline
\hspace{4mm}S &A&T&T&G&C&T&A&C\\
\hline
  \hspace{4mm} i  &0&1&2&3&4&5&6&7\\
 \hline
 $RANK_{i}$ &${\perp}$&5&4&${\perp}$&2&3&${\perp}$&1\\
\hline
\end{tabular}
\end{table}\\\\
STEP 2 search Non-sample Suffixes\\
  Given each Nonsample suffix $S_{i}
\in S_{B_{0}}$
with the pair $(t_{i}$
,$rank(S_{i}+1)$. It is so obvious that : for $i, j \in B_{0} $,
$S_{i}\le S(j) \leftrightharpoons (t_{i},rank(S_{i+1}))\le (t_{j},rank(S_{j+1}))$. For example, $S_{6}$$\le$$S_{0}$$\le$$S_{3}$, because$ (A,1)\le (A,5)\le (G,2))$\\\\
STEP 3 merge those of the two segments\\
Combining two segments into one string should be obey the rules :
\\ Let $S_{i}\in S_{C}$ with $S_{j}\in S_{B_{0}}$,\\\\
$i\in B_{1}$ :  $S_{i}\le S(j) \leftrightharpoons(t_{i},rank(S_{i+1}))\le (t_{j},rank(S_{j+1}))$\\\\
$i\in B_{2} $ :  $S_{i}\le S(j)\leftrightharpoons(t_{i},t_{i+1},rank(S_{i+1}))\le(t_{j},t_{j+1},rank(S_{j+1}))$.
Finally we can obtain sorted sequence in Table 3.
\begin{table}[!htbp]\centering\normalsize
\caption{RANK OF SORTED SAMPLE SUFFIX}
\begin{tabular}{|l|l|l|l|l|l|l|l|l|}
\hline
\hspace{4mm}S &A&T&T&G&C&T&A&C\\
\hline
  \hspace{4mm} i  &0&1&2&3&4&5&6&7\\
 \hline
 $rank_{S_{i}}$ &2&8&7&5&4&6&1&3\\
\hline
\end{tabular}
\end{table}\\\\
 After construction of suffix array, we should find the longest overlap prefix of two sequences we assigned. The lcp between two suffixes is the minimum of the lcps of all pairs of adjacent
suffixes between them on the Pos array \cite{kasai2001linear}. That is :\\\\
$lcp(S_{pos_{i}},S_{pos_{y}})=\min_{i< k\le j} \lbrace lcp(S_{pos_{k}},S_{pos_{k+1}}) \rbrace$.\\\\

 Range-Minimum-Query-Problem is to preprocess an array so that the position of the minimum element between two specified indices can be obtained efficiently \cite{gusfield1997algorithms}
 \cite{yang2010efficient}. First we give a definition to the Range Minimum Query (RMQ) : \\

Given an array S[1...n] of elements from a totally ordered set (with order relation "${\le}$", $RMQ_{S}(i,j)$ returns the index of a smallest element in A[i, j], i.e., $RMQ_{S_{i,j}}$=$argmin_{k{\in}(i,..j)}$$\lbrace S[k] \rbrace$. Illustrate in table.\\

 So LCP problem can be successfully transformed into RMQ problem. The method was first proposed and presented by Bender and Farach-Colton. we present in this paper that Berkman and Vishkin algorithm which is combined with $\pm$1RMQ is other case of RMQ problem. Disadvantages of the algorithm is waste of the large space, but it maintains efficiency of query process in $<O(n),O(1)>$ to complete.
\begin{table}[!htbp]\centering\normalsize
\caption{SUFFIX INDEX TABLE OF $S{\lbrace}ATTGCTAC{\rbrace}$}
\begin{tabular}{|l|l|l|l|l|l|l|l|l|}
\hline
\hspace{4mm}S &A&T&T&G&C&T&A&C\\
\hline
  INDEX  &0&1&2&3&4&5&6&7\\
 \hline
\hspace{3mm}SA &6&0&7&4&3&5&2&1\\
\hline
\hspace{2mm} lcp &0&1&0&1&0&0&1&1\\
\hline
\end{tabular}
\end{table}\\\\

 This algorithm can be divided into two parts :\\\\

STEP 1  RMQ problem can be converted to LCA problem(least common ancestor): first build cartesian tree store the input sequence of A , build the complexity of the cartesian tree is O (n). \\

 STEP 2  Transform LCA problems into $\pm$1RMQ : based on DFS search of the tree by Euler path (Euler Tour), set up three arrays  E, L and R. The E and L size are $2*n-1$, elements of E is label value for each node of the cartesian tree (is actually index of A), elements of L is corresponding to the depth of the node of euler path. R stores the positions of the each node queried for the first time.\\\\

 If want to query LCA(u,v), it is actually equivalent to solving RMQ problem of the array of L. Due to do RMQ can be transformed into $\pm$RMQ, it also can use the $\pm$1RMQ algorithm in $<(n), O(1)>$ time.
 \section{accelerate contruction of lcp table and solve largest overlap region problem with cuda}

   Sorting is a common problem in bioinformatics, like quick sort. Merge sort etc. Quick sort algorithm not only guarantee the computing process on
hardware which is shortage of  atomic operations but also utilize geometry shaders, however is generally slightly slower and quick sort is testified as efficiency algorithm. In DC3 algorithm,
the construction process of the sample suffix array and the non-sample suffix both use radix sort to built array. To many a certain keys problem solving by comparison-base algorithm, Radix sort is relatively an efficient and low-cost method. This sorting algorithm contains b processes which process j-th digits of the keys in order from the least to the most digit. Previously, many compute scientists have developed many variant algorithm, especially using bitonic sort (Govindaraju et al. 2006, Zachmann 2006). In our paper, we present a efficient radix sort to implement DC3 \cite{greb2006gpu} \cite{farach1997optimal}.\\\\

\begin{figure}[H]
\centering
\graphicspath{}
\includegraphics[width=2.5in]{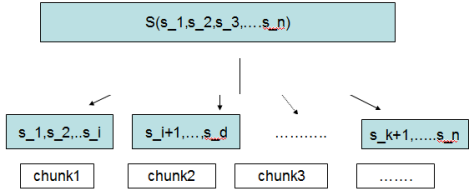}
 \DeclareGraphicsExtensions.
\caption{chunks of divided S}
\label{fig 17}
\end{figure}

\begin{figure}[H]
\centering
\graphicspath{}
\includegraphics[width=2.5in]{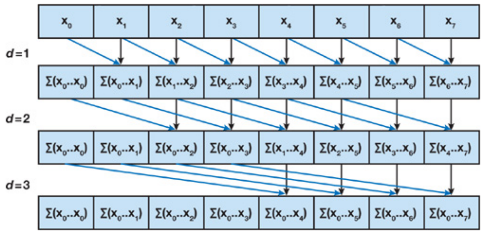}
 \DeclareGraphicsExtensions.
\caption{scan process}
\label{fig 16}
\end{figure}

\begin{figure}[H]
\centering
\graphicspath{}
\includegraphics[width=2.5in]{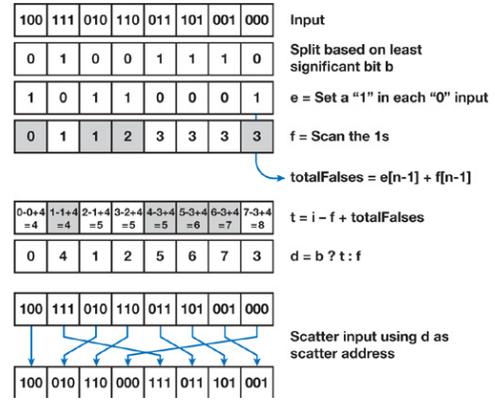}
 \DeclareGraphicsExtensions.
\caption{DC3 process}
\label{fig 15}
\end{figure}

 This algorithm presents each chunk of subsequence is sorted by radix sort. a grid of thread blocks cooperatively sort chunks in a parallel way. Due to a lack of enough size of shared memory, it makes difficult for single multiprocessors to allocate more space to chunks of input array. Given a string $S=\lbrace s_{1}s_{2}s_{3}..s_{n}\rbrace$, divide S into every single chunk like $chunk_{1}=\lbrace s_{1}s_{2}..s_{i}\rbrace$, $chunk_{2}=\lbrace s_{i+1}s_{i+2}..s_{d}\rbrace$, ..$chunk_{..}=\lbrace s_{..},..,s_{n}\rbrace$.illustrate in Figure.\ref{fig 17}.
 Taka a number which has m-bits keys as an example, it need to take m steps to finish radix sort algorithm. Thus every step of process requires us to do scan the bit table. It is notable that split primitive is significant for all parallel process to achieve an efficient and satisfied  performance. After that we should organize each sort keys which are gained by split into a certain keys list, and each value represent the least significant bit that is either "1" or "0". But there are some tricks to produce new reversed the table list that all "0" sorted key are replaced by "1"  and vice versa. Finally we need to formulate a rule to compute the destination address of each input number. For example, $chunk1=\lbrace001,100,111,101,110\rbrace$, let array L store the least significant bit b, so $L=\lbrace1,0,1,1,0\rbrace$. Define E is a array, and for $i \ in\lbrace0\le i\le n\rbrace$, $E[i]=L[i] \otimes  1 $. And we can obtain the $E=\lbrace 0,1,0,0,1\rbrace$. Scan E to obtain F. Illustrate in Figure.\ref{fig 16}.
  $F=\lbrace0,1,1,2,3 \rbrace$. Process illustrate in the Figure.\ref{fig 16}. Using array F[n-1] and E[n-1] to obtain totalFalse, $totalfalse=E[n-1]+F[n-1]$. Final index of each element in S can be computed by formula : $index=b?(i-f+totalFalse):F $. All the process follow Figure.\ref{fig 15} \cite{nguyen2007gpu}.

\begin{algorithm}[H]         
\caption{ KERNEL FUNCTION FOR PARALLEL RADIX SORT IN DC3 ALGORITHM  }             
\label{alg:SA}                  
\begin{algorithmic}[1]                
\WHILE {each $i < 32$}
  \STATE  $row=blockIdx.x*twidth+threadIdx.x$;
  \STATE  $col=blockIdx.y*twidth+threadIdx.y$;
  \STATE   $id=row*width+col$;
  \STATE  $ set the index for each threads$
  \STATE  $b[id]=( S[id] \gg i)\&1$;
  \STATE   ${//}$ record i{\_}th bit of each element
  \STATE  ${\_\_}syncthreads()$;
   \STATE  $  e[id]=b[id]\otimes1$;
   \STATE  $ {\_\_}syncthreads()$;
  \STATE $temp[id] = e[id]$;
\STATE ${\_\_}syncthreads()$;
\FORALL{$oFF=1 $ , \textbf{and},$oFF<n$}
\STATE ${//}$scan the 1s table to calculate prefix sum of each element
\STATE $oFF\ll 1$;
\IF{$id \le oFF$}
\STATE $temp[id]=temp[id-oFF]+temp[id]$;
\ELSE
\STATE$ temp[id] = temp[id]$;
\ENDIF
\STATE ${\_\_}synctreads()$;
\ENDFOR
\STATE $f[id]=temp[id]$;
\STATE  ${\_\_}syncthreads()$;
\STATE $ tof=e[n-1]+f[n-1]$
 \STATE $t[id]=id-f[id]+tof$;
 \STATE${\_\_}syncthreads()$;
 \STATE$d[id]=b[id]?t[id]:f[id]$;
 \STATE ${//}$array d store the index of the sorted S
 \STATE${\_\_}syncthreads()$;
\STATE $OutS[d[id]]=S[id]$;
\STATE $S=OutS$;
  \ENDWHILE

\end{algorithmic}
\end{algorithm}
When all assigned thread blocks chunks finished, each chunk of sorted numbers should be repeatedly combined into a new sorted list and each new sorted list is allocated to a coalescent chunk until all chunks finish merge sort process. The pseudocode is depicted
in Alg.\ref{alg:SA} in the kernel function cudaDC3RS. In the kernel function id is
a private register with respect to a thread respectively, which denote as thread index.

Before the process depicted as Alg.\ref{alg:SA}, we should divide whole elements of S into sample suffix and Nonsample suffix. here we just present a single process of sample suffix. Let S to be sample suffix. Follow the Alg.\ref{alg:SA} all single block of subsequence are organized
into a string array S ,so that the exclusive thread can
be mapped into S[id]. Then each subsequence is
executed respectively so as to the indexed structure enable
all threads to string matching simultaneously by means of
radix sort algorithm supported by GPU.

\section{result}
In this section we compare the sequential performance with parallel performance of CUDA implementation. Both algorithms are implemented using Microsoft Visual Studio 2010 combined
with CUDA version 5.5 for the parallel implementation. The proposed algorithms are carried out in a Intel(R)
Core(TM) i3-3110K quad-core running at 2.40GHZ with
2.0GB RAM. The used CUDA driver and runtime version are
both 5.0, NVIDIA GeForce 610M GPU(kepler architecture)
which has a total of 48 streaming multiprocessors operating
at a clock rate of 900 MHZ. \\\\

   Noncoding DNA sequences from NCBI Nucleotide are to
evaluate the performance of the previously described algorithms. The reference sequences corresponds to 3 groups: Homo sapiens chromosome Y noncoding region downstream from the DAZ gene of which accession is AH011747.1 GI: 21929706, Caenorhabditis elegans Bristol N2 genomic chromosome of which accession is
 BX284603.4 GI: 449020129 and Homo sapiens isolate NA19204 noncoding region T1419 genomic sequence of which accession is  GQ846167.1 GI: 260538176. Several groups of query sequences were used to experimental test,
each one consists of a different number of query sequences,
ranging from 512 to 2097152 query sequences. It is worth noticing that the programmer writes a kernel and organize its execution in a grid of thread blocks, Each block is assigned to a Streaming Multiprocessor (SM). Once assigned it cannot migrate to another SM. Each SM splits its own blocks into Warps (currently with a maximum size of 32 threads). All the threads in a warp executes concurrently on the resources of the SM. The actual execution of a thread is performed by the CUDA Cores contained in the SM. It is a very important process for mapping noncoding DNA sequences into the grid. Illustrate in Table.5.\\\\

\begin{figure}[H]
\centering
\graphicspath{}
\includegraphics[width=2.5in]{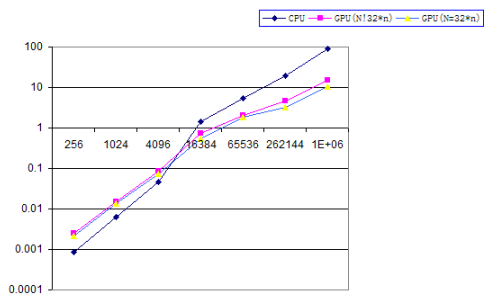}
 \DeclareGraphicsExtensions.
\caption{performance of CPU and kernel execution of GPU}
\label{fig 19}
\end{figure}

\begin{figure}[H]
\centering
\graphicspath{}
\includegraphics[width=2.5in]{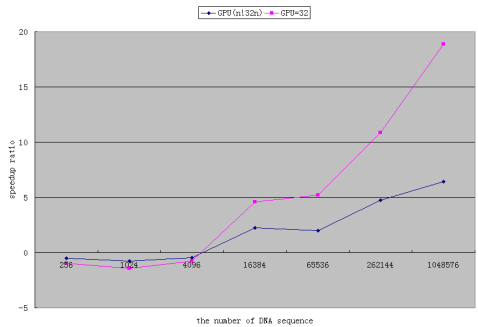}
 \DeclareGraphicsExtensions.
\caption{average speedup ratio between GPU and CPU }
\label{fig 20}
\end{figure}

 Accept for implementation of finding the longest overlap region of noncoding DNA on CUDA, CPU programme of the algorithm also is executed for sake of evaluating  the different efficiency between the series programme and parallel programme. And the series parts execute on Intel(R) Core(TM) i3-3110M CPU. The results is illustrated in Figure.\ref{fig 19}. The result shows DC3 radix sort algorithm on GPU
  in O(n) time to construct suffix array compared with execution on CPU. And it is easy to find that the performance on GPU is not better than CPU execution when the number of the noncoding DNA sequences are lower than 65536. After more the number of noncoding DNA sequences, performance improvement on the GPU of algorithm is much satisfied. It also observed
that efficiency of  performance is growing with
significantly growing the numbers of sequences. Illustrate in Figure.\ref{fig 20}. Unlike what happened in CPU implementations,
tile optimization partition local data into shared memory, due to the achieved performance in accordance with memory accesses, reduce the access of global memory.

\begin{table}[!htbp]\centering\tiny
\caption{DC3 radix sort based on CPU and GPU}
\begin{tabular}{|l|l|l|l|l|l|l|l|}
\hline
\hspace{4mm} {TYPE / NUMBER}&256&1024&4096&16384&65536&262144&1048576\\
\hline
\hspace{4mm}  G1CPU  &0.001&0.0062&0.047&1.425&5.325&19.375&98.253\\
 \hline
\hspace{3mm}$G1GPU!32n$ &0.002&0.015&0.085&0.741&2.012&4.524&9.1012\\
\hline
\hspace{3mm}$G1GPU=32n$ &0.002&0.014&0.072&0.531&1.842&3.254&8.335\\
\hline
\hspace{4mm}  G2CPU  &0.001&0.0068&0.0051&1.615&5.827&22.941&98.453\\
 \hline
 \hspace{3mm}$G2GPU!32n$ &0.002&0.016&0.087&0.6981&1.962&5.014&8.232\\
\hline
\hspace{3mm}$G2GPU=32n$ &0.002&0.014&0.088&0.513&1.901&3.044&7.963\\
\hline
\hspace{4mm}  G3CPU  &0.001&0.0067&0.049&1.577&5.553&21.531&95.616\\
 \hline
 \hspace{3mm}$G3GPU!32n$ &0.002&0.019&0.087&0.720&2.232&5.271&7.816\\
\hline
\hspace{3mm}$G3GPU=32n$ &0.002&0.015&0.081&0.681&1.822&3.116&6.735\\
\hline

\end{tabular}
\end{table}

\begin{table}[!htbp]\centering\tiny
\caption{Speedup rating of different execution}
\begin{tabular}{|l|l|l|l|l|l|l|l|}
\hline
\hspace{4mm} {TYPE / SPEED RATIO}&256&1024&4096&16384&65536&262144&1048576\\
 \hline
\hspace{3mm}$G1GPU!32n$ &-0.5&-0.75&-0.47&2.25&2.61&4.28&10.79\\
\hline
\hspace{3mm}$G1GPU=32n$ &-0.48&-0.67&-0.31&2.32&2.89&5.96&11.79\\
\hline
\hspace{3mm}$G2GPU!32n$ &-0.5&-0.71&-0.49&2.75&2.96&4.57&11.85\\
\hline
\hspace{3mm}$G2GPU=32n$ &-0.48&-0.61&-0.28&3.15&3.07&7.52&12.37\\
\hline
\hspace{3mm}$G3GPU!32n$ &-0.5&-0.69&-0.47&2.25&2.39&4.08&12.24\\
\hline
\hspace{3mm}$G3GPU=32n$ &-0.48&-0.67&-0.31&2.32&3.04&6.91&14.20\\
\hline
\end{tabular}
\end{table}

From the obtained
results, it can be observed that runtime of CPU
were consistently higher than implemented DC3 radix sort, it shows solving the longest overlap region problem with respect
to suffix array construction using GPU even is efficient method
to high performance bioinformatics applications.
\section{conclusion}
  This paper presents a new method to solve problem of the longest overlap region of Noncoding DNA sequence using GPU hardware and modified algorithm. Related experiment were thoroughly compared using two different executional ways: multicore (i3-3110M) and many-core (NVIDIA GeForce
GTX610M GPU). \\\\
  These observations reveal that massive data applied to the longest overlap problem is time-consuming job. Nevertheless,
parallel programming on GPU achieves the improvement in rate
accelerating making solving DNA sequence more
precise and more faster so that can meet our requirement in more
important projects.\\

\bibliography{RM}
\bibliographystyle{IEEEtran}

\end{document}